\documentclass[letter]{myaa}
\usepackage{graphicx}
\usepackage{txfonts}
\usepackage{natbib}
\usepackage[a4paper,breaklinks]{hyperref}
\idline{1}{1}
\begin{document}
\def\teff{$T\rm_{eff}$ }
\def\kms {\,$\mathrm{km\, s^{-1}}$ }
\def\kmss {\,$\mathrm{km\, s^{-1}}$}
\def\ms {$\mathrm{m\, s^{-1}}$ }

\newcommand{\Teff}{\ensuremath{T_\mathrm{eff}}}
\newcommand{\g}{\ensuremath{g}}
\newcommand{\gf}{\ensuremath{gf}}
\newcommand{\loggf}{\ensuremath{\log\gf}}
\newcommand{\glog}{\ensuremath{\log\g}}
\newcommand{\pun}[1]{\,#1}
\newcommand{\cobold}{\ensuremath{\mathrm{CO}^5\mathrm{BOLD}}}
\newcommand{\linfor}{Linfor3D}
\newcommand{\xx}{\ensuremath{\mathrm{1D}_{\mathrm{LHD}}}}
\newcommand{\punms}{\mbox{\rm\,m\,s$^{-1}$}}
\newcommand{\punkms}{\mbox{\rm\,km\,s$^{-1}$}}
\newcommand{\abuhe}{\mbox{Y}}
\newcommand{\grav}{\ensuremath{g}}
\newcommand{\mlp}{\ensuremath{\alpha_{\mathrm{MLT}}}}
\newcommand{\mlpcm}{\ensuremath{\alpha_{\mathrm{CMT}}}}
\newcommand{\moh}{\ensuremath{[\mathrm{M/H}]}}
\newcommand{\senv}{\ensuremath{\mathrm{s}_{\mathrm{env}}}}
\newcommand{\shelio}{\ensuremath{\mathrm{s}_{\mathrm{helio}}}}
\newcommand{\smin}{\ensuremath{\mathrm{s}_{\mathrm{min}}}}
\newcommand{\spun}{\ensuremath{\mathrm{s}_0}}
\newcommand{\sstar}{\ensuremath{\mathrm{s}^\ast}}
\newcommand{\tauross}{\ensuremath{\tau_{\mathrm{ross}}}}
\newcommand{\ttaurelation}{\mbox{T$(\tau$)-relation}}
\newcommand{\Ysurf}{\ensuremath{\mathrm{Y}_{\mathrm{surf}}}}
\newcommand{\mD}{\ensuremath{\left\langle\mathrm{3D}\right\rangle}}

\newcommand{\draftflag}{false}

\newcommand{\beq}{\begin{equation}}
\newcommand{\eeq}{\end{equation}}
\newcommand{\pdx}[2]{\frac{\partial #1}{\partial #2}}
\newcommand{\pdf}[2]{\frac{\partial}{\partial #2}\left( #1 \right)}

\newcommand{\var}[1]{{\ensuremath{\sigma^2_{#1}}}}
\newcommand{\sig}[1]{{\ensuremath{\sigma_{#1}}}}
\newcommand{\cov}[2]{{\ensuremath{\mathrm{C}\left[#1,#2\right]}}}
\newcommand{\xtmean}[1]{\ensuremath{\left\langle #1\right\rangle}}

\newcommand{\eref}[1]{\mbox{(\ref{#1})}}

\newcommand{\Vact}{\ensuremath{\nabla}}
\newcommand{\Vad}{\ensuremath{\nabla_{\mathrm{ad}}}}
\newcommand{\Veddy}{\ensuremath{\nabla_{\mathrm{e}}}}
\newcommand{\Vrad}{\ensuremath{\nabla_{\mathrm{rad}}}}
\newcommand{\Vraddiff}{\ensuremath{\nabla_{\mathrm{rad,diff}}}}
\newcommand{\cp}{\ensuremath{c_{\mathrm{p}}}}
\newcommand{\taueddy}{\ensuremath{\tau_{\mathrm{e}}}}
\newcommand{\vconv}{\ensuremath{v_{\mathrm{c}}}}
\newcommand{\Fconv}{\ensuremath{F_{\mathrm{c}}}}
\newcommand{\lmix}{\ensuremath{\Lambda}}
\newcommand{\Hp}{\ensuremath{H_{\mathrm{P}}}}
\newcommand{\Hptop}{\ensuremath{H_{\mathrm{P,top}}}}
\newcommand{\COBOLD}{{\sc CO$^5$BOLD}}

\newcommand{\changed}{}

\newcommand{\I}{\ensuremath{I}}
\newcommand{\Irot}{\ensuremath{\tilde{I}}}
\newcommand{\F}{\ensuremath{F}}
\newcommand{\Frot}{\ensuremath{\tilde{F}}}
\newcommand{\vsini}{\ensuremath{V\sin(i)}}
\newcommand{\vvsini}{\ensuremath{V^2\sin^2(i)}}
\newcommand{\vsinimu}{\ensuremath{\tilde{v}}}
\newcommand{\rotint}{\ensuremath{\int^{+\vsinimu}_{-\vsinimu}\!\!d\xi\,}}
\newcommand{\imu}{\ensuremath{m}}
\newcommand{\imupone}{\ensuremath{{m+1}}}
\newcommand{\nmu}{\ensuremath{N_\mu}}
\newcommand{\msum}[1]{\ensuremath{\sum_{#1=1}^{\nmu}}}
\newcommand{\wmu}{\ensuremath{w_\imu}}

\newcommand{\tchar}{\ensuremath{t_\mathrm{c}}}
\newcommand{\Nt}{\ensuremath{N_\mathrm{t}}}

\title{The solar photospheric abundance of phosphorus: }
\subtitle{results from \cobold\ 3D model atmospheres}

\author{
E. Caffau     \inst{1},
M. Steffen   \inst{2},
L. Sbordone  \inst{1,3},
H.-G. Ludwig  \inst{1,3}, and
P. Bonifacio  \inst{1,3,4}
}
\institute{GEPI, Observatoire de Paris, CNRS, Universit\'e Paris Diderot; Place
Jules Janssen 92190
Meudon, France
\and
Astrophysikalisches Institut Potsdam, An der Sternwarte 16, D-14482 Potsdam, Germany
\and
CIFIST Marie Curie Excellence Team
\and
Istituto Nazionale di Astrofisica,
Osservatorio Astronomico di Trieste,  Via Tiepolo 11,
I-34143 Trieste, Italy
}
\authorrunning{Caffau et al.}
\titlerunning{Solar phosphorus abundance}
\offprints{Elisabetta.Caffau@obspm.fr}
\date{Received ...; Accepted ...}

\abstract
{}
{We determine the solar abundance of phosphorus
using \cobold\ 3D hydrodynamic model atmospheres.}
{High resolution, high signal-to-noise solar spectra
of the \ion{P}{i} lines of Multiplet 1 at 1051-1068\pun{nm}
are compared to line formation computations performed
on a \cobold\ solar model atmosphere.}
{We find A(P)=$5.46\pm 0.04$, in good agreement with previous
analysis based on 1D model atmospheres, 
due to the fact that the \ion{P}{i} lines
of Mult. 1 are little affected by 3D effects.
We cannot confirm an earlier claim by other authors of a downward revision of the solar P
abundance by 0.1\,dex employing a 3D model atmosphere. 

Concerning other stars, we found modest ($<0.1$\,dex) 3D abundance corrections for P among four F-dwarf model
atmospheres of different metallicity, being largest at lowest metallicity.  
}
{We conclude that 3D abundance corrections are generally rather small for the
\ion{P}{i} lines studied in this work. They are marginally relevant for
metal-poor stars, but may be neglected in the Sun.}
\maketitle


\section{Introduction}

Phosphorus is an element whose nucleosynthesis
is not clearly understood and whose abundance is
very poorly known outside the solar system.
It has a single stable isotope
$^{31}$P, and
its most likely site of production
are  carbon and neon burning shells
in the late stages of the evolution
of massive stars, which end up as Type II SNe.
The production mechanism is probably
through neutron capture, as it is for
the  parent nuclei $^{29}$Si and $^{30}$Si.
According to  \citet{woosley95} 
there is no significant P production
during the explosive phases.

Like for other odd-Z elements (Na, Al)
one expects that the P abundance should
be proportional to the neutron
excess $\eta$\footnote{as defined
in \citet{arnet71}:
$\eta = (n_{\rm n} - n_{\rm p})/(n_{\rm n} + n_{\rm p})$, where
$n_{\rm n}$ represents the total number of neutrons per gram, both
free and bound in nuclei, and $n_{\rm p}$ the corresponding number for
protons}. As metallicity decreases the
neutron excess decreases and these odd-Z elements should
decrease more rapidly than neighbouring even-Z elements.
One can therefore expect decreasing [$^{31}$P/$^{30}$Si] 
or [$^{31}$P/$^{12}$Mg]
with decreasing metallicity.
Such a behaviour is in fact observed for Na
\citep{andrievsky07}, however at low metallicity
[Na/Mg] seems to reach a plateau at about [Na/Mg]$\sim -0.5$.
One could expect a similar behaviour for P.

Phosphorus abundances in stars have so far been determined for
object classes not
apt to study the chemical
evolution: 
chemically peculiar stars (e.g. \citealt{castelli97,fremat}, for
a review of observations in Hg-Mn stars see \citealt{takada} and references
therein),
horizontal branch stars \citep{behr,bonifacio95,BS76}, sub-dwarf B type stars
\citep{ohl00,baschek},
Wolf-Rayet stars \citep{marcolino}, PG 1159 stars \citep{jahn}, and
white dwarfs \citep{chayer,dobbie,vennes}.
Such stars are not useful to trace the chemical
evolution 
because either 
they are chemically peculiar
or they are evolved and may
have changed their original chemical composition. 
A few measurements exist
in O super-giants \citep{bouret,crowther02} and B type stars \citep{tobin}.
\citet{kato} measured P in Procyon,
\citet{sion} in the dwarf nova VW Hydri.

The abundance of P in F, G and K stars, which could be derived
from the high excitation IR \ion{P}{i} lines of Mult. 1 
at 1051--1068\,nm
has never been explored, due to 
the few high resolution near IR spectrographs available. 
With the advent of
the CRIRES spectrograph at the VLT, the situation is
likely to change and it will be possible to study the
chemical evolution of P in the Galaxy using these long-lived stars  
as tracers.
In this perspective 
it is interesting, as a reference, to revisit the solar P abundance
in the light of the recent advances of 3D hydrodynamical
model atmospheres. There are several determinations of the solar
P abundance in the literature. Given  the difficulty of 
this observation, it is not surprising that P was absent
from the seminal work of \citet{russell}. 
\citet{lambert68} measured  an abundance 
A(P)\footnote{A(P) = $\log(N_{\rm P}/N_{\rm H})+12$} 
= 5.43, using 
13 lines of \ion{P}{i}, of which four belong to Mult. 1 .
The oscillator strength they used for these lines are very close to the 
\citet{biemont73} values (see Table \ref{gflog}).
\citet{lambert78} measured A(P) $=5.45\pm 0.03$,
which is essentially the photospheric abundance adopted in the
authoritative 
\citet{anders89} compilation
(A(P)$=5.45\pm 0.04$).
Based on their {\it ab initio} computed oscillator strengths \citet{biemont94} 
derived A(P) $=5.45\pm 0.06$. With computed oscillator 
strengths, corrected to match measured energy level lifetimes, 
\citet{berzinsh97} derived  $5.49\pm 0.04$.
In their compilation
\citet{grevesse98} adopt A(P) 
$=5.45\pm 0.04$.
As can be seen, all determinations
of the solar photospheric phosphorus abundance 
are in agreement, within the stated uncertainties.
In disagreement with earlier work,
\citet{sunabboasp} obtained a significantly
lower
value of A(P)$=5.36\pm 0.04$ using a 3D solar model atmosphere
computed with the Stein \& Nordlund code (see \citealt{nordlund97}).
This result might be immediately interpreted as indication that 3D abundance
corrections lead to a lower
solar phosphorus abundance by $\sim 0.1$\,dex, but other factors,
such as the choice of \loggf\ values, can also explain this low A(P).
The A(P) determinations found in literature are summarised in 
Table~\ref{pilit}.

In the present paper we use a solar 3D \cobold\ model to re-assess the solar
phosphorus abundance.  For comparison we also consider 1D solar models.
Beyond the Sun, we further discuss abundance corrections
obtained for a number of F-dwarf models of varying metallicity.


\section{Atomic data}

We consider five IR \ion{P}{i} lines of Mult. 1
(see Table \ref{irpline}).
Several determinations of the
oscillator strengths for this multiplet
are available in the literature (see Table \ref{gflog}). Generally, all values are
rather close, differences are smaller than 0.1\,dex for any line except the
1051.2\pun{nm} line for which the maximum difference is 0.18\,dex.
In this work we adopted the values computed by \citet{berzinsh97},
corrected semi-empirically,
which, in our opinion, are the
best currently available.
Had we chosen another set of values,
with the notable exception of the \citet{berzinsh97}
{\it ab initio} values, the difference in the derived phosphorus
abundance would be 0.05\,dex at most.

\begin{table}
\caption{Infra-red phosphorus lines considered in this work.}
\label{irpline}
\begin{center}
\begin{tabular}{rrrrr}
\hline
\noalign{\smallskip}
 Wavelength & Transition & E$_{\rm low}$ & \loggf  \\
 {[nm]}     &            & {[eV]}\\
\noalign{\smallskip}
\hline
\noalign{\smallskip}
 1051.1584 & 4s$^4$P$_{1/2}$--4p$^4$D$^0_{3/2}$ & 6.94 & --0.13 \\
 1052.9522 & 4s$^4$P$_{3/2}$--4p$^4$D$^0_{5/2}$ & 6.95 &   0.24 \\
 1058.1569 & 4s$^4$P$_{5/2}$--4p$^4$D$^0_{7/2}$ & 6.99 &   0.45 \\
 1059.6900 & 4s$^4$P$_{1/2}$--4p$^4$D$^0_{1/2}$ & 6.94 & --0.21 \\
 1068.1402 & 4s$^4$P$_{3/2}$--4p$^4$D$^0_{3/2}$ & 6.95 & --0.19 \\
\noalign{\smallskip}
\hline
\end{tabular}
\end{center}
\end{table}


\section{Models}

Our analysis is based on a 3D model atmosphere
computed with the \cobold\ code \citep{freytag02,wedemeyer04}.
In addition to the \cobold\ hydrodynamical
simulation we used several 1D models.
More details about the employed models can be found
in \citet{caffau2007} and \citet{zolfito}.  
In the following we will refer to the 
1D model derived by a
horizontal and temporal averaging of 
the 3D \cobold\ model as the \mD\ model. The 
reference 1D model for the computation of the
3D abundance corrections is a related (same \Teff, \glog, and chemical
composition) hydrostatic
1D model atmosphere computed with the LHD code,
hereafter \xx\ model.
LHD is a Lagrangian 1D hydrodynamical model
atmosphere code, which employs the same micro-physics as \cobold. Convection
is treated within the mixing-length approach.
In the LHD model computation, as in any 1D theoretical atmosphere model,
at least one free parameter describing the treatment of convection must be
decided upon: we use a mixing-length parameter of $\alpha$=1.25 in the solar 
case, a value of $\alpha$=1.8 in [M/H]=--2.0, and $\alpha$=1.0 in [M/H]=--3.0 
models. We 
employ the formulation of \citet{mihalas78} for mixing-length theory, and we
neglect turbulent pressure in the momentum equation.
We also considered the solar ATLAS\,9 model 
computed by Fiorella Castelli
\footnote{http://wwwuser.oats.inaf.it/castelli/sun/ap00t5777g44377k1asp.dat}
and the empirical Holweger-M\"uller solar model \citep{hhsunmod, hmsunmod}.

The spectrum synthesis codes employed
are \linfor\footnote{http://www.aip.de/~mst/Linfor3D/linfor\_3D\_manual.pdf}
for all models,
and SYNTHE 
\citep{1993KurCD..18.....K,2005MSAIS...8...14K},
in its Linux version \citep{2004MSAIS...5...93S,2005MSAIS...8...61S},
for the fitting procedure with 1D or \mD\ models. 
The advantage of using SYNTHE, with respect to the present 
version of \linfor, is that it can handle easily a large
number of lines, thus allowing to take into account numerous
weak blends.

The 3D \cobold\ solar model we used is the same we describe
in \citet{zolfito}.
It covers a time interval of 6000\pun{s}, represented by 25 snapshots;
this covers about 10 convective turn-over
time scales, and 20 periods of the 5 minute oscillations which are
present in the 3D model as acoustic modes of the computational box.

Like in our study of sulphur \citep[see][]{zolfito},
we define as ``3D abundance correction'' the difference in the abundance
derived from the full 3D model and the related \xx\ model,
in the sense A(3D)\,-\,A(\xx), both synthesised
with \linfor.
We consider also  the difference in the abundance derived
from the 3D model and from the \mD\ model.
Since the 3D and \mD\ models have, by construction, the same 
mean temperature structure, this allows to single out the 
effects due to the horizontal temperature fluctuations.


\section{Data}

As observational data we 
used two high-resolution, high signal-to-noise ratio spectra of 
disk-centre solar intensity:  that of \citet{neckelobs} (hereafter
referred to as the  ``Neckel intensity
spectrum'') and that of Delbouille, Roland, Brault, \&\ Testerman (1981)
(hereafter referred to as the ``Delbouille intensity spectrum''
\footnote{http://bass2000.obspm.fr/solaru\_spect.php}).
We also used the solar flux spectrum of \citet{neckelobs} and the solar flux
spectrum of \citet{Ksun}\footnote{http://kurucz.harvard.edu/sun.html},
referred to as the ``Kurucz flux spectrum''.


\section{Data analysis}

We measured the equivalent width (EW) of the lines
using the IRAF task {\tt splot}.
We then computed the phosphorus abundance for the
solar photosphere from the curve of growth of the line in question calculated with \linfor.
As a confirmation of our results we fitted all observed line profiles
with synthetic profiles. 
We found a good agreement 
in comparison to the abundances derived from the EWs.
The remaining differences we found 
are probably related to the difference in the continuum opacities 
used by SYNTHE with respect to the ones used by \linfor.

\begin{figure}
\resizebox{7.5cm}{!}{\includegraphics[clip=true,angle=0]{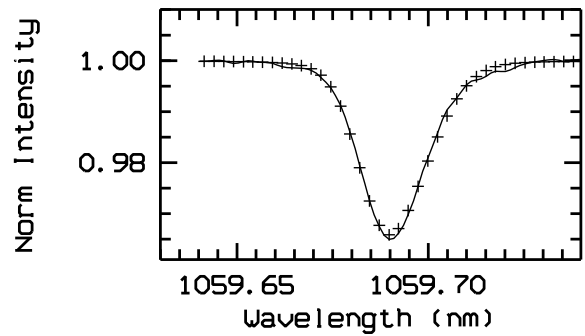}}
\caption{The observed disk-centre Neckel intensity spectrum 
(solid line) is
plotted over the fit (crosses) obtained using a grid of synthetic
spectra based on the \cobold\ model computed
with \linfor.}
\label{fitint}
\end{figure}

The line profile fitting was done using
a code, described in \citet{zolfo05}, which performs
a $\chi ^2$ minimisation of the deviation between synthetic profiles
and the observed spectrum.
Figure~\ref{fitint} shows the fit of the solar intensity spectrum obtained
from a grid of synthetic spectra synthesised with \linfor\ and
based on the 3D \cobold\ model. The close agreement between observed and 
synthetic line profile indicates that the non-thermal (turbulent) line
broadening is very well represented by the hydrodynamical velocity field
of the \cobold\ simulation.
 
\section{Results and discussion}

\subsection{The solar P abundance}

The different sets of \loggf\ values for the selected 
\ion{P}{i} lines are very close,
so that the derived  solar A(P) is quite insensitive 
to this choice.
The solar phosphorus abundance varies
by 0.05\,dex, from the highest value obtained with the data set of
\citet{biemont94} to the lower one obtained using \citet{biemont73},
the corrected value of \citet{berzinsh97}, or the one from \citet{KP}.
A lower value of the solar phosphorus abundance (0.09\,dex below the
one obtained using the \citet{biemont94} oscillator strength)
can be obtained using the non-corrected values from \citet{berzinsh97}.

With the (corrected) \loggf\ of \citet{berzinsh97} we obtain
A(P)=$5.443\pm 0.058$ for the flux spectra considering the whole sample of five lines,
A(P)=$5.426\pm 0.064$ for the intensity spectra at disk-centre.
The line at 1068.1\pun{nm} yields an abundance which is 1.6 $\sigma$ below the mean value,
while the other four lines are within one $\sigma$. If we remove
this line from the computation, the standard deviation drops by a factor of two and we obtain
A(P)=$5.467\pm 0.029,~5.450\pm 0.042$ for flux and intensity spectra, respectively.
The differences in the A(P) determination from intensity and flux
spectra could be due to NLTE corrections which usually are different in
intensity and flux. Unfortunately, there are no NLTE computations available
for these lines. 
Our recommended  solar photospheric abundance is A(P)=$5.458\pm 0.036$,
very close to all previous 1D determinations, however
0.1\,dex higher than the value found by \citet{sunabboasp} using a 3D model.

The result of  \citet{sunabboasp} is difficult to interpret since, according
to our model,
the 3D corrections for the \ion{P}{i} lines in the Sun are small.
\citet{sunabboasp} do not provide detailed information on the lines
used or of their adopted EWs.
They adopted the \loggf\ of \citet{berzinsh97},
however they do not specify if they used the ``corrected'' or 
{\em ab initio} values. If they used the {\em ab initio}
values the difference is easily understood, otherwise we must
conclude that the difference is due either to a difference 
between their 3D simulation and the \cobold\ one, or due to
a difference in the adopted EWs. We point out here that
our measured EWs are in quite good agreement with both
those of \citet{lambert68}, 
\citet{lambert78}
and with those of \citet{biemont94}

\begin{figure}
\resizebox{7.5cm}{!}{\includegraphics[clip=true,angle=0]{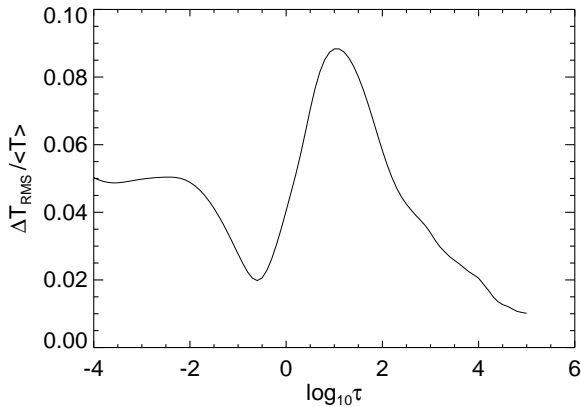}}
\caption{Relative horizontal RMS temperature fluctuations
(on surfaces of equal Rosseland optical depth) in the solar \cobold\ model.
}
\label{tscat}
\end{figure}

A detailed summary of the solar phosphorus abundances we derived from 
individual lines, using different model atmospheres, in given in 
Table~\ref{sunzolfo}.
Since the considered lines are not truly weak, the 3D corrections
are unfortunately sensitive to the adopted micro-turbulence, and 
hence must be interpreted with care.
The 3D corrections due to horizontal temperature fluctuations only, 
indicated by the 3D-\mD\ difference, are slightly positive throughout.
In all lines, the 3D-\mD\ difference for intensity spectra is systematically
higher than for flux spectra, by roughly 0.02\,dex.
This is presumably due to the fact that the intensity spectra
originate from somewhat deeper layers where horizontal fluctuations 
are larger. We note that the phosphorus lines are formed mostly in the 
range $-0.5<\log_{10}\tau _{\rm ross}<0$,
where the horizontal temperature fluctuations increase with increasing 
optical depth (see Fig. \ref{tscat}). 

The 3D-\mD\ difference increases with the excitation potential of the
line's lower level. In fact, the 3D-\mD\ correction is rather small,
not exceeding +0.030\,dex for $\xi$=1.0~\kmss, and +0.045\,dex for
$\xi$=1.5~\kmss, which we consider as an upper limit.

The 3D-\xx\ corrections are systematically larger than the 
3D-\mD\ difference (by +0.01 to 0.02\,dex), but of the same order of 
magnitude. From that we can conclude that the solar \mD\ and \xx\
temperature profiles are very similar. 

\subsection{3D effects on P for other stars.}

As shown above,
we obtain only mild 3D effects in our solar phosphorus abundance analysis.
They are comparable to the standard deviation (as listed in col.\ (9) of
Table~\ref{sunzolfo}) for the weak lines, and a factor four larger than the 
standard deviation for the strongest line, and hence the
3D effects are hardly relevant. However, 3D effects should become stronger 
for metal-poor stars where the temperature profiles differ more strongly 
between \mD\ and \xx\ models than for solar metallicity models
\citep{asp2005}.
Very little work has been dedicated to 3D effects on the phosphorus abundance,
and none for metal-poor stellar models.
To check the behaviour of the
3D abundance corrections, we theoretically investigated for a
few available 3D \cobold\ models the phosphorus 3D-\xx\ relation in 
the resulting synthetic flux spectra.

We find, for three metal-poor models (6300/4.5/--2.0\footnote{indicating
\Teff/logg/[M/H]}; 5900/4.5/--3.0; 6500/4.5/--3.0), that the difference in 
the mean temperature profile is the only factor which contributes to the 3D
correction. In the metal-poor models, as in the solar model, the maximum
contribution to the equivalent widths of the \ion{P}{i} lines comes from a
well defined region close to the continuum forming layers.
Even though the optical depth range contributing significantly is somewhat 
broader in the metal-poor models than in the solar model, the \mD\ and \xx\ 
models are usually rather close in temperature in this region, and so the 
effects of the different mean temperature structure are not very pronounced
for the considered \ion{P}{i} lines. The 3D correction related to the 
difference in the mean temperature structure of the 3D and \xx\ model amounts 
to $\approx$ +0.05\,dex for the --2.0 metallicity model, and to 
$\approx$ +0.1\,dex for the most metal-poor model.

For a 3D \cobold\ model of Procyon (6500/4.0/0.0) the difference in
the temperature profile of the \mD\ and \xx\ model contributes to more
than half of the total 3D correction.  In Procyon, \ion{P}{i} lines
are formed at shallower optical depths than in the Sun by about
$\Delta\log\tau\approx 0.3$.  In this range the \xx\ model is cooler
than the 3D \cobold\ one.  The related abundance correction depends on
the excitation potential (strength) of the line, and for the strongest
line it is still less than +0.030\,dex.  The contribution to the 3D
correction related to the horizontal temperature fluctuations ranges
from +0.022 to +0.045\,dex, increasing with the excitation potential
(strength) of the line. In the range where the line is formed, the
horizontal RMS temperature fluctuations in Procyon are of the order of
5\% to be compared to 1\% in the metal-poor models. The total 3D 
correction is thus less than +0.07\,dex for all lines.


\section{Conclusions}

Using the \cobold\ solar granulation model, we have determined
the solar 3D LTE phosphorus abundance to be
A(P)=$5.46\pm 0.04$, which compares very well to all previous
determinations obtained using 1D models. This can be explained by
the fact that the \ion{P}{i}
lines of Mult. 1 appear to be rather insensitive
to granulation effects, at least in the parameter interval
explored by us (F and G dwarfs of metallicity from solar to --3.0).
These lines can thus be useful to investigate the chemical evolution
of phosphorus in the Galactic disk and in moderately 
metal-poor environments.
Below [P/H]=--1.0, however, the lines have EWs smaller than 0.5\pun{pm},
becoming exceedingly difficult to observe.

For the \ion{P}{i} lines studied in this work, we find small (Sun) to 
moderate (Procyon, metal-poor stars) 3D abundance corrections. The sign 
of the corrections is found to be \emph{positive} in all cases,
meaning that the abundances resulting from a 3D analysis are larger
than those obtained from a 1D model.
Note that this behaviour is consistent with the results
obtained by \citet{ms02} for \ion{S}{i}, which has almost
the same ionisation potential as \ion{P}{i} (see their Fig.~8).
In general, however, the magnitude and sign of the 3D effects depend on
the properties of the absorbing ion (ionisation potential, the excitation 
potential of the lower level, line strength and wavelength), and on the 
thermodynamic structure of the stellar atmosphere. The results obtained
in this work for \ion{P}{i} must therefore not be generalised to other
elements and other types of stars.


\begin{acknowledgements}
  The authors L.S., H.-G.L., P.B. acknowledge financial
  support from EU contract MEXT-CT-2004-014265 (CIFIST).
  We acknowledge use of supercomputing centre CINECA,
  which has granted us time to compute part of the hydrodynamical
  models used in this investigation, through the INAF-CINECA
  agreement 2006,2007.
\end{acknowledgements}

\bibliographystyle{aa}

\begin{thebibliography}{}

\bibitem[Aller(1949)]{1949ApJ...109..244A} Aller, L.~H.\ 1949, \apj, 109,
244

\bibitem[Anders \& Grevesse(1989)]{anders89} Anders, E., \& 
Grevesse, N.\ 1989, \gca, 53, 197 

\bibitem[Andrievsky et al.(2007)]{andrievsky07} Andrievsky, S.~M., 
Spite, M., Korotin, S.~A., Spite, F., Bonifacio, P., Cayrel, R., Hill, V., 
\& Fran{\c c}ois, P.\ 2007, \aap, 464, 1081

\bibitem[Arnett(1971)]{arnet71} Arnett, W.~D.\ 1971, \apj, 166, 
153 

\bibitem[Asplund(2005)]{asp2005} Asplund, M.\ 2005, \araa, 43, 
481 

\bibitem[Asplund et al.(2005)]{sunabboasp} Asplund, M., Grevesse, 
N., \& Sauval, A.~J.\ 2005, ASP Conf.~Ser.~336: Cosmic Abundances as 
Records of Stellar Evolution and Nucleosynthesis, 336, 25 

\bibitem[Baschek et al.(1982)]{baschek} Baschek, B., Scholz,
M., Kudritzki, R.~P., \& Simon, K.~P.\ 1982, \aap, 108, 387

\bibitem[Baschek \&\ Sargent (1976)]{BS76} Baschek, B., \&
Sargent, A.~I.\ 1976, \aap, 53, 47

\bibitem[Behr et al. (1999)]{behr} Behr, B.~B., Cohen, J.~G.,
McCarthy, J.~K., \&\ Djorgovski, S.~G.\ 1999, \apjl, 517, L135

\bibitem[Berzinsh et al. (1997)]{berzinsh97} Berzinsh, U., 
Svanberg, S., \&\ Biemont, E.\ 1997, \aap, 326, 412 

\bibitem[Biemont et al.(1994)]{biemont94} Biemont, E., Martin, 
F., Quinet, P., \& Zeippen, C.~J.\ 1994, \aap, 283, 339 

\bibitem[Biemont \&\ Grevesse (1973)]{biemont73} Biemont, E. and
Grevesse, N.\ 1973, At. Data Nucl. Data Tables 12, 217

\bibitem[Bonifacio et al.(1995)]{bonifacio95} Bonifacio, P., 
Castelli, F., \& Hack, M.\ 1995, \aaps, 110, 441 

\bibitem[Bouret et al.(2005)]{bouret} Bouret, J.-C., Lanz, T., 
\& Hillier, D.~J.\ 2005, \aap, 438, 301 

\bibitem[Caffau \& Ludwig (2007)]{zolfito} Caffau, E., \& 
Ludwig, H.-G.\ 2007, \aap, 467, L11

\bibitem[Caffau et al. (2005)]{zolfo05} Caffau, E., Bonifacio, 
P., Faraggiana, R., Fran{\c c}ois, P., Gratton, R.~G., \& Barbieri, M.\ 
2005, \aap, 441, 533 
 
\bibitem[Caffau et al.(2007)]{caffau2007} Caffau, E., Faraggiana, 
R., Bonifacio, P., Ludwig, H.-G., \& Steffen, M.\ 2007, \aap, 470, 699


\bibitem[Castelli et al.(1997)]{castelli97} Castelli, F., 
Parthasarathy, M., \& Hack, M.\ 1997, \aap, 321, 254


\bibitem[Chayer et al.(2005)]{chayer} Chayer, P., Vennes, S., 
Dupuis, J., \& Kruk, J.~W.\ 2005, \apjl, 630, L169 

\bibitem[Crowther et al. (2002)]{crowther02} Crowther, P.~A., 
Hillier, D.~J., Evans, C.~J., Fullerton, A.~W., De Marco, O., \& Willis, 
A.~J.\ 2002, \apj, 579, 774 

\bibitem[Dobbie et al. (2005)]{dobbie} Dobbie, P.~D., Barstow, 
M.~A., Hubeny, I., Holberg, J.~B., Burleigh, M.~R., \& Forbes, A.~E.\ 2005, 
\mnras, 363, 763 

\bibitem[Fremat \&\ Houziaux (1997)]{fremat} Fremat, Y., \&
Houziaux, L.\ 1997, \aap, 320, 580

\bibitem[Freytag et al. (2002)]{freytag02} Freytag, B., Steffen, 
M., \& Dorch, B.\ 2002, AN, 323, 213

\bibitem[Garcia \&\ Levato (1986)]{1986ApL....25....1G} Garcia, Z.~L., \&
Levato, H.\ 1986, \aplett, 25, 1

\bibitem[Grevesse \&\ Sauval (1998)]{grevesse98} Grevesse, N., \& 
Sauval, A.~J.\ 1998, Space Science Reviews, 85, 161 

\bibitem[Holweger (1967)]{hhsunmod} Holweger, H.\ 1967, 
Zeitschrift f{\"u}r Astrophysik, 65, 365 

\bibitem[Holweger \&\ M\"uller (1974)]{hmsunmod} Holweger, H., \& 
Mueller, E.~A.\ 1974, \solphys, 39, 19 

\bibitem[Jahn et al.(2007)]{jahn} Jahn, D., Rauch, T., 
Reiff, E., Werner, K., Kruk, J.~W., \& Herwig, F.\ 2007, \aap, 462, 281 

\bibitem[Kato et al.(1996)]{kato} Kato, K.-I., Watanabe, Y.,
\& Sadakane, K.\ 1996, \pasj, 48, 601

\bibitem[Kaufmann \&\ Schonberger (1977)]{1977A&A....57..169K} Kaufmann,
J.~P., \& Schonberger, D.\ 1977, \aap, 57, 169

\bibitem[{{Kurucz}(1993{\natexlab{b}})}]{1993KurCD..18.....K}
{Kurucz}, R. 1993{\natexlab{b}}, SYNTHE Spectrum Synthesis Programs and Line
Data.~Kurucz CD-ROM No.~18.~Cambridge, Mass.: Smithsonian Astrophysical
Observatory, 1993., 18

\bibitem[{{Kurucz}(2005{\natexlab{a}})}]{2005MSAIS...8...14K}
{Kurucz}, R.~L. 2005{\natexlab{a}}, MSAIS, 8, 14

\bibitem[Kurucz(2005)]{Ksun} Kurucz, R.~L.\ 2005, MSAIS, 
 8, 189

\bibitem[Kurucz \& Peytremann (1975)]{KP}
{Kurucz}, R.~L., \&\ {Peytremann}, E.\ 1975, SAO Special Report, 362 

\bibitem[Lambert \&\ Warner(1968)]{lambert68} Lambert, D.~L., \& 
Warner, B.\ 1968, \mnras, 138, 181 

\bibitem[Lambert \&\ Luck(1978)]{lambert78} Lambert, D.~L., \& 
Luck, R.~E.\ 1978, \mnras, 183, 79 

\bibitem[Marcolino et al.(2007)]{marcolino} Marcolino, W.~L.~F., 
Hillier, D.~J., de Araujo, F.~X., \& Pereira, C.~B.\ 2007, \apj, 654, 1068 

\bibitem[Mihalas (1978)]{mihalas78} Mihalas, D.\ 1978, San 
Francisco, W.~H.~Freeman and Co., 1978.~650 p.

\bibitem[Neckel \&\ Labs (1984)]{neckelobs} Neckel, H., \& Labs, 
D.\ 1984, \solphys, 90, 205

\bibitem[Nordlund \& Stein(1997)]{nordlund97} Nordlund, A., \& 
Stein, R.\ 1997, SCORe'96 : Solar Convection and Oscillations and their 
Relationship, 225, 79

\bibitem[Ohl et al. (2000)]{ohl00} Ohl, R.~G., Chayer, P., \& 
Moos, H.~W.\ 2000, \apjl, 538, L95

\bibitem[Russell (1929)]{russell} Russell, H.~N.\ 1929, \apj,
70, 11

\bibitem[{{Sbordone} (2005)}]{2005MSAIS...8...61S}
{Sbordone}, L. 2005, MSAIS, 8, 61

\bibitem[{{Sbordone} {et~al.} (2004){Sbordone}, {Bonifacio}, {Castelli}, \&
  {Kurucz}}]{2004MSAIS...5...93S}
{Sbordone}, L., {Bonifacio}, P., {Castelli}, F., \& {Kurucz}, R.~L. 2004,
  MSAIS, 5, 93

\bibitem[Seligman \&\ Aller (1970)]{1970Ap&SS...9..461S} Seligman, C.~E., \&
Aller, L.~H.\ 1970, \apss, 9, 461

\bibitem[Sion et al. (1997)]{sion} Sion, E.~M., Cheng, F.~H.,
Sparks, W.~M., Szkody, P., Huang, M., \& Hubeny, I.\ 1997, \apjl, 480, L17

\bibitem[Steffen \&\ Holweger (2002)]{ms02} Steffen, M., \& 
Holweger, H.\ 2002, \aap, 387, 258

\bibitem[Takada-Hidai (1991)]{takada} Takada-Hidai, M.\ 1991, 
IAU Symp.~145: Evolution of Stars: the Photospheric Abundance Connection, 
145, 137 

\bibitem[Tobin \&\ Kaufmann (1984)]{tobin} Tobin, W., \&
Kaufmann, J.~P.\ 1984, \mnras, 207, 369

\bibitem[Vennes et al. (1996)]{vennes} Vennes, S., Chayer, P.,
Hurwitz, M., \& Bowyer, S.\ 1996, \apj, 468, 898

\bibitem[Wedemeyer et al. (2004)]{wedemeyer04} Wedemeyer, S., 
Freytag, B., Steffen, M., Ludwig, H.-G., \& Holweger, H.\ 2004, \aap, 414, 
1121 

\bibitem[Woosley \& Weaver (1995)]{woosley95} Woosley, S.~E., \& 
Weaver, T.~A.\ 1995, \apjs, 101, 181 


\end{thebibliography}

\Online

\begin{table}
\caption{Spectroscopic determinations of the solar photospheric phosphorus abundance.}
\label{pilit}
\begin{center}
\begin{tabular}{ccl}
\hline
\noalign{\smallskip}
A(P) & $\sigma$ & Reference\\
\noalign{\smallskip}
\hline
\noalign{\smallskip}
5.43 &      &\citet{lambert68}\\
5.45 & 0.03 &\citet{lambert78}\\
5.45 & 0.04 &\citet{anders89}\\
5.45 & 0.06 &\citet{biemont94}\\
5.49 & 0.04 &\citet{berzinsh97}\\
5.45 & 0.04 &\citet{grevesse98}\\
5.36 & 0.04 &\citet{sunabboasp}\\
5.46 & 0.04 & this work\\
\noalign{\smallskip}
\hline
\end{tabular}
\end{center}
\end{table}

\begin{table}
\caption{Oscillator strength for the IR phosphorus lines
(wavelengths in nm) available in the literature.}
\label{gflog}
\begin{center}
{\scriptsize
\begin{tabular}{rrrrrl}
\hline
\noalign{\smallskip}
\multicolumn{5}{c}{\loggf} & Reference\\
 1051.2 & 1053.0 & 1058.2 & 1059.7 & 1068.1 \\
\noalign{\smallskip}
\hline
\noalign{\smallskip}
--0.21  & 0.20   &  0.48  & --0.21 & --0.10 &\citet{KP}\\
--0.23  & 0.18   &        & --0.24 & --0.12 &NIST\\
--0.06  & 0.31   &  0.52  & --0.14 & --0.12 &\citet{berzinsh97} \\
--0.13  & 0.24   &  0.45  & --0.21 & --0.19 &\citet{berzinsh97} corr.\\
--0.10  & 0.27   &  0.49  & --0.17 & --0.15 &\citet{biemont94}\\
--0.24  & 0.17   &  0.45  & --0.24 & --0.13 &\citet{biemont73}\\
\noalign{\smallskip}
\hline
\end{tabular}
}
\end{center}
(NIST: \href{http://physics.nist.gov/PhysRefData/ADS}{http://physics.nist.gov/PhysRefData/ADS})
\end{table}

\begin{table*}
\begin{center}
\caption{Solar phosphorus abundances from the various observed spectra,
adopting the (corrected) \loggf\ of \citet{berzinsh97}.
\label{sunzolfo}}
{
\begin{tabular}{rrrrrrrrrrrrr}
\hline
\noalign{\smallskip}
 Spectrum &Wave      &  EW    & \multicolumn{5}{c}{A(P) from EW} & $\sigma$ & \multicolumn{2}{c}{3D-\xx} & \multicolumn{2}{c}{3D-\mD}\\
          &{[nm]}    & {[pm]} & {[dex]}  &  {[dex]} & {[dex]}  & {[dex]}  & {[dex]}  & {[dex]}  & {[dex]}  & {[dex]}  & {[dex]} & {[dex]}\\
          &          &        & 3D   &  \multicolumn{2}{c}{ATLAS} & \multicolumn{2}{c}{HM} \\
          &        &      &   & 1.0  &   1.5  &   1.0  &   1.5  &   & 1.0  &   1.5  &   1.0  &   1.5 \\  
 (1)      &  (2)   & (3)  &  (4)   &  (5)   &  (6)   &  (7)   &  (8)   &  (9)   &  (10)  &  (11)  & (12)  & (13)\\
\noalign{\smallskip}
\hline
\noalign{\smallskip}
 KF & 1051.2&  0.674 &   5.482&   5.476&   5.471&   5.480&   5.475&   0.015&   0.015&   0.019&   0.005&   0.010\\
 NF & 1051.2&  0.734 &   5.523&   5.517&   5.512&   5.521&   5.516&   0.013&   0.015&   0.020&   0.005&   0.010\\
 NI & 1051.2&  0.811 &   5.502&   5.476&   5.473&   5.498&   5.493&   0.012&   0.028&   0.032&   0.022&   0.026\\
 DI & 1051.2&  0.822 &   5.508&   5.483&   5.479&   5.504&   5.500&   0.012&   0.028&   0.032&   0.022&   0.027\\
\noalign{\smallskip}
\hline
\noalign{\smallskip}
 KF & 1053.0&  1.330 &   5.431&   5.421&   5.412&   5.427&   5.417&   0.010&   0.019&   0.028&   0.005&   0.015\\
 NF & 1053.0&  1.340 &   5.435&   5.424&   5.416&   5.431&   5.421&   0.010&   0.019&   0.028&   0.005&   0.015\\
 NI & 1053.0&  1.520 &   5.425&   5.392&   5.382&   5.417&   5.408&   0.010&   0.036&   0.044&   0.025&   0.034\\
 DI & 1053.0&  1.550 &   5.436&   5.402&   5.392&   5.427&   5.418&   0.010&   0.036&   0.044&   0.025&   0.034\\
\noalign{\smallskip}
\hline
\noalign{\smallskip}
 KF & 1058.2&  2.280 &   5.480&   5.463&   5.448&   5.473&   5.456&   0.012&   0.027&   0.042&   0.006&   0.023\\
 NF & 1058.2&  2.120 &   5.436&   5.421&   5.406&   5.429&   5.414&   0.013&   0.025&   0.039&   0.005&   0.021\\
 NI & 1058.2&  2.330 &   5.407&   5.363&   5.346&   5.393&   5.379&   0.012&   0.047&   0.059&   0.029&   0.043\\
 DI & 1058.2&  2.330 &   5.407&   5.363&   5.346&   5.393&   5.379&   0.012&   0.047&   0.059&   0.029&   0.043\\
\noalign{\smallskip}
\hline
\noalign{\smallskip}
 KF & 1059.7&  0.684 &   5.482&   5.475&   5.471&   5.480&   5.475&   0.014&   0.015&   0.019&   0.005&   0.010\\
 NF & 1059.7&  0.660 &   5.464&   5.458&   5.454&   5.462&   5.458&   0.014&   0.015&   0.019&   0.005&   0.010\\
 NI & 1059.7&  0.748 &   5.456&   5.431&   5.427&   5.452&   5.448&   0.012&   0.027&   0.030&   0.022&   0.026\\
 DI & 1059.7&  0.751 &   5.458&   5.433&   5.429&   5.454&   5.450&   0.012&   0.027&   0.031&   0.022&   0.026\\
\noalign{\smallskip}
\hline
\noalign{\smallskip}
 KF & 1068.1&  0.670 &   5.370&   5.364&   5.360&   5.368&   5.364&   0.014&   0.015&   0.019&   0.005&   0.010\\
 NF & 1068.1&  0.617 &   5.330&   5.324&   5.320&   5.328&   5.324&   0.015&   0.015&   0.019&   0.006&   0.010\\
 NI & 1068.1&  0.699 &   5.322&   5.297&   5.294&   5.318&   5.314&   0.013&   0.027&   0.030&   0.022&   0.026\\
 DI & 1068.1&  0.731 &   5.343&   5.319&   5.315&   5.340&   5.336&   0.013&   0.027&   0.030&   0.022&   0.026\\
\noalign{\smallskip}
\hline
\noalign{\smallskip}
\end{tabular}
}
\end{center}
Col.~(1) spectrum identification: DI: Delbouille intensity, NI: Neckel 
intensity, NF: Neckel flux, KF: Kurucz flux.
Col. (2) wavelength of the line.
Col.~(3) measured equivalent width.
Col.~(4) phosphorus abundance, A(P), according to the \cobold\ 3D model.
Cols.~(5)-(8) provide A(P) from 1D models, odd numbered cols. correspond
to a micro-turbulence~$\xi$ of 1.0\kms, even numbered cols. to $\xi = 1.5$\kms.
Col.~(9) uncertainty in A(P) due to the uncertainty in the measured EWs.
Col.~(10)-(13) provide 3D corrections, even numbered cols. for
$\xi = 1.0$\kmss, and odd numbered cols. for 1.5\kmss, respectively.
\end{table*}

\end{document}